\documentclass[aps, pra, preprint, groupedaddress, amsfonts, longbibliography,
               amsmath, amssymb, showpacs, nofootinbib]{revtex4-1}

\usepackage{microtype}
\usepackage{graphicx}
\usepackage{epstopdf}
\usepackage[utf8]{inputenc}
\usepackage[T1]{fontenc}
\usepackage[usenames,dvipsnames]{xcolor}
\usepackage{hyperref}

\usepackage{color}% \textcolor{#1}{#2}

\begin{document}
\title{Electron capture and ionization cross-section calculations for proton collisions from 
methane and the DNA and RNA nucleobases}

\author{Hans J\"urgen L\"udde}
\email[]{luedde@itp.uni-frankfurt.de}
\affiliation{Frankfurt Institute for Advanced Studies (FIAS), D-60438 Frankfurt, Germany} 

\author{Marko Horbatsch}
\email[]{marko@yorku.ca}
\affiliation{Department of Physics and Astronomy, York University, Toronto, Ontario M3J 1P3, Canada}

\author{Tom Kirchner}  
\email[]{tomk@yorku.ca}
\affiliation{Department of Physics and Astronomy, York University, Toronto, Ontario M3J 1P3, Canada}
\date{\today}
\begin{abstract}
Net ionization and net capture cross-section calculations are presented for proton collisions from
methane molecules and the DNA/RNA nucleobases adenine, cytosine, guanine, thymine, and uracil.
We use the recently introduced independent-atom-model pixel counting method to calculate these
cross sections in the 10 keV to 10 MeV impact energy range and compare them
with results obtained from the simpler additivity rule, a previously used
complete-neglect-of-differential-overlap method, 
and with experimental data and previous calculations where available.
It is found that all theoretical results agree reasonably well at high energies, but deviate
significantly in the low-to-intermediate energy range. In particular, the pixel counting method which takes
the geometrical overlap of atomic cross sections into account is the only calculation that is
able to describe the measurements for capture in proton-methane collisions down to 10 keV impact energy.
For the nucleobases it also yields a significantly smaller cross section in this region than the
other models. New measurements are urgently required to test this prediction.
\end{abstract}

\maketitle
\section{Introduction}
\label{intro}
Collisions involving complex molecules pose a significant challenge to theory 
owing to their large number of degrees of freedom and their multi-center geometry.
A convenient framework for a simplified discussion is offered by the independent atom model (IAM)
according to which certain properties and quantities of a molecule are {\it constitutive}, i.e.,
can be obtained by adding up atomic contributions. 
When applied to collisions, the simplest version of the IAM consists in adding up the cross sections of the
atomic constituents of a given molecule in order to obtain the molecular cross section.
This procedure is usually referred to as the
additivity rule (AR). It goes back to Bragg~\cite{bragg05} 
and was first studied in a more systematic fashion in the 1950s
for electron-impact ionization of medium-sized molecules~\cite{otvos56}. 
The IAM-AR has since been applied to many
electron- and ion-impact collision systems, typically with good success at high impact energies,
while larger discrepancies with experimental data have been found in the low-to-intermediate energy region.
This has been taken as an indication that molecular structure effects gain importance
with decreasing collision energy~\cite{galassi00}.

Accordingly, extensions and alternatives have been considered, such as ARs with weight factors
(see, e.g., \cite{Blanco2003} and the review article~\cite{Deutsch00} and references therein)
and a complete-neglect-of-differential-overlap (CNDO) 
approach~(see, e.g., \cite{galassi00, Champion12, quinto18}.
The latter starts from the assumption that a molecular net cross section 
can be expressed as a sum of partial cross sections for
all initially occupied molecular orbitals (MOs) 
and then approximates 
those partial cross sections 
as linear combinations of atomic-orbital (AO) specific cross sections 
with weight factors that
are obtained from a Mulliken population analysis~\cite{mulliken55}.
In this way, the CNDO approach takes molecular structure information into
account to some extent. However,
CNDO cross sections do not depend on the orientation of the molecule relative to the projectile beam direction, i.e.,
they are to be interpreted as orientation-averaged quantities. The same is, of course, true
for the simple IAM-AR. 
Moreover, none of these methods account for the ro-vibrational motion of the molecule,
but this is uncritical for the impact energy ranges considered.

We have recently introduced an amended AR with orientation-dependent weight factors~\cite{hjl16, hjl18}. The weights
are obtained from a geometrical interpretation of a molecular cross section as an effective area made
up of overlapping atomic cross sections.
The atomic cross sections
are calculated using a first-principles-based
method with ground-state density-functional-theory (DFT) potentials~\cite{tom98},
and the effective area 
encountered by the impinging projectile is computed 
using a pixel counting
method. The latter step is repeated for a large number of orientations so that the orientation average
can be compared with
experimental data for randomly oriented molecules.
We refer to this model as IAM-PCM~\cite{hjl18}. 

Previous applications of the IAM-PCM to proton collisions from medium-sized molecules such as H$_2$O and
from larger compounds such as water clusters and a variety of biomolecules 
have shown promising results~\cite{hjl16, hjl18, hjl19}.
At high impact energies where the simple IAM-AR works well, the overlap effects are small and the IAM-PCM
cross sections agree with the IAM-AR results. Toward lower energies they deviate and tend to be in 
better agreement with the (scarce) experimental data, i.e., our results suggest that the 
IAM-PCM represents an improvement compared to the IAM-AR. 

The purpose of the present work is to further establish the 
IAM-PCM as a viable tool for net ionization and capture cross section calculations for ion-molecule collision
systems.
Our focus is proton 
impact on methane molecules (CH$_4$) and the five DNA and RNA nucleobases 
adenine (C$_5$H$_5$N$_5$), cytosine (C$_4$H$_5$N$_3$O), 
guanine (C$_5$H$_5$N$_5$O), thymine (C$_5$H$_6$N$_2$O$_2$) and uracil (C$_4$H$_4$N$_2$O$_2$),
for which
some experimental data and previous theoretical results, mostly obtained within the CNDO approach, are
available for comparison. We also 
present CNDO results based on our own first-principles atomic cross section calculations and
demonstrate that they disagree with the IAM-PCM predictions in regions in which overlap effects are
strong. The discrepancies are most pronounced for electron capture at relatively low
impact energies where the projectile
speed is similar to or smaller than the average orbital speed of the molecular valence electrons.
We call for experimental efforts to validate (or refute) the IAM-PCM predictions for the
nucleobases.

The paper is organized as follows. We briefly summarize the IAM-PCM and contrast it
with the CNDO approach in Sect.~\ref{sec:theory}. Results are presented in 
Sect.~\ref{sec:results}. 
We start with a look at
net ionization and capture in the proton-methane collision system to illustrate
a few general trends and then discuss our
results for proton collisions from the DNA/RNA nucleobases.
The paper ends with a few concluding remarks in Sect.~\ref{sec:conclusions}.
Atomic units, characterized by $\hbar=m_e=e=4\pi\epsilon_0=1$, are used unless otherwise stated.

\section{Theoretical models}
\label{sec:theory}
The IAM-PCM amounts to representing a net cross section for a molecular target
at projectile energy $E$ and for process $x$, where
$x$ stands for capture ($x={\rm cap}$) or ionization ($x={\rm ion}$), 
as a weighted sum
of atomic cross sections
\begin{equation}
 \sigma_{\rm IAM-PCM}^{{\rm net} \, x}(E,\alpha,\beta,\gamma) = 
  \sum_{j=1}^N s_j^x(E,\alpha,\beta,\gamma)  \sigma_j^{{\rm net} \, x} (E) 
\label{eq:pcm1}
\end{equation}
with weight factors $0\le s_j^x\le 1$ for the $N$ atoms that make up the molecule, and
the Euler angles
$\alpha,\beta,\gamma$ 
which characterize the orientation of the molecule relative to the projectile beam axis.
The atomic cross sections are calculated in a DFT-inspired framework in the semiclassical
approximation with straight-line projectile trajectories~\cite{tom98}. The
initially occupied atomic orbitals (AOs) are propagated in a mean-field potential
composed of a (time-dependent) Coulombic projectile potential and an atomic target
ground-state potential at the exchange-only level of DFT, i.e., correlation and time-dependent screening
and exchange effects are neglected. The propagation is carried out
using the coupled-channel two-center basis
generator method (TC-BGM)~\cite{tcbgm} and the net cross sections for ionization
and capture are obtained by summation of
the corresponding orbital-specific transition probabilities and integration over the
impact-parameter plane.

The weights in Eq.~(\ref{eq:pcm1}) are obtained by picturing the atomic 
cross sections as circular disks around the equilibrium nuclear positions in the molecule
and with radii
$r_j(E) = [\sigma_j^{{\rm net} \, x}(E)/\pi]^{1/2}$ in a plane perpendicular to the ion beam axis.
The combined area of overlapping circles that is ``visible'' to the  impinging projectile is interpreted as the
molecular cross section and computed by a pixel counting method~\cite{hjl18}.
The procedure is carried out for a large number of Euler angle triples to obtain an orientation-averaged
cross section that can be compared with experimental data for randomly-oriented molecules.

Keeping in mind that the atomic cross sections are composed of AO-specific contributions
we can write for the orientation average 
\begin{equation}
 \bar\sigma_{\rm IAM-PCM}^{{\rm net} \, x}(E) = 
  \sum_{j} \bar s_j^x(E) \sum_{k} n_{k,j}^{\rm ao} \sigma_{k,j}^{{\rm ao} \, x} (E) ,
\label{eq:pcm2}
\end{equation}
where the sum over $k$ includes all AOs on the $j$th atom with nonzero occupation numbers
$n_{k,j}^{\rm ao}$ and the $\bar s_j^x$ are orientation-averaged weight factors.
Since the occupation numbers and the orbital-specific cross sections are the same
for each atom of a certain species,
Eq.~(\ref{eq:pcm2}) can be cast into the
simpler form
\begin{equation}
 \bar\sigma_{\rm IAM-PCM}^{{\rm net} \, x}(E) = 
  \sum_i  \eta_i^x(E)  \sigma_i^{{\rm ao} \, x} (E) ,
\label{eq:pcm3}
\end{equation}
where the index $i$ enumerates the occupied AOs of {\it different} atomic species only 
and $\eta_i^x$ is a coefficient that is composed
of occupation numbers and orientation-averaged weight factors. 
In the limiting case of zero overlap in which $\bar s_j^x=1$ for all $j$ and $x$ one obtains
\begin{equation}
 \sigma_{\rm IAM-AR}^{{\rm net} \, x}(E) = 
  \sum_i n_i^{\rm ao} \sigma_{i}^{{\rm ao} \, x} (E) ,
\label{eq:tcs-ar}
\end{equation}
where $ n_i^{\rm ao} $ is the {\it total} occupation number of the $i$th AO in the molecule.
Note that by construction the IAM-AR result (\ref{eq:tcs-ar}) represents an upper bound for the IAM-PCM cross section (\ref{eq:pcm3}).

Let us compare these equations with the CNDO approach, which starts from the assumption that the net cross section
is composed of MO-specific contributions
\begin{equation}
 \sigma_{\rm CNDO}^{{\rm net} \, x}(E) = 
  \sum_{l} n_l^{\rm mo} \sigma_{l}^{{\rm mo} \, x} (E) 
\label{eq:tcs-mo}
\end{equation}
with occupation numbers $n_l^{\rm mo}$.
In a second step the MO-specific cross sections $\sigma_{l}^{{\rm mo} \, x}$ are expressed as linear
combinations of AO-specific cross sections. If the latter are independent of the MOs to which they contribute
the CNDO net cross section can be written as 
\begin{equation}
 \sigma_{\rm CNDO}^{{\rm net} \, x}(E) = 
  \sum_i  \xi_{i} \, \sigma_{i}^{{\rm ao} \, x} (E) ,
\label{eq:tcs-cndo}
\end{equation}
where $\xi_{i}$ is the (fractional) {\it gross}
population in the $i$th AO due to {\it all} initially occupied MOs~\cite{mulliken55} 
and the sum includes all AOs that are (partially) populated by at least one of the MOs. 
If one restricts the population analysis to the minimal atomic basis used in the expansion of the MOs, 
the index ranges in Eqs.~(\ref{eq:tcs-cndo}), (\ref{eq:tcs-ar}), and (\ref{eq:pcm3}) are the same.
Hence, the only difference between the CNDO approach, the IAM-AR and the IAM-PCM is the nature of the coefficients
in these linear combinations of AO-specific cross sections. In the IAM-AR they are simply the atomic
occupation numbers. In the CNDO approach they include molecular structure information via the
Mulliken analysis, while in the
IAM-PCM they depend on the impact energy and the process under study, because the atomic cross
section overlaps do.

A caveat of the foregoing analysis is that in the previously reported CNDO calculations the AO-specific cross
sections are not completely independent of the MOs to which they contribute. 
In those works, the AO-specific calculations were
carried out in some variant of the first-order Born approximation or a distorted-wave model, all of which involve the use
of effective target charges in the construction of the final (continuum or bound projectile) states of interest.
These effective charges were determined using {\it molecular} binding energies in Bohr's energy formula,
thereby introducing an MO dependence into the AO-specific cross sections~(see, e.g.,~\cite{galassi10, quinto18}).
By contrast, there is no room for such a choice in the coupled-channel TC-BGM and, accordingly, 
our own CNDO calculations, reported here, do satisfy Eq.~(\ref{eq:tcs-cndo}).

\section{Results and discussion}
\label{sec:results}
Before we look at the DNA/RNA nucleobases let us exemplify the different theoretical models discussed
above for the proton-methane collision system as a test case. 
Table~\ref{tab:ch4} lists the total atomic occupation numbers and the fractional gross populations 
obtained from
a Mulliken analysis for CH$_4$~\cite{hoffmann63} together with the AO binding energies~\cite{aashamer78}.
Given that the gross population of the most weakly bound orbital 
$\xi_{{\rm C}(2p)}=3.399$ is significantly larger than the AO occupation number $n_{{\rm C}(2p)}^{\rm ao}=2$ 
and ionization tends to increase with
decreasing $\varepsilon_i$,
one can expect the total net ionization cross section calculated within
the CNDO approach to be larger than its IAM-AR counterpart.

\begin{figure}
\begin{center}
\resizebox{0.6\textwidth}{!}{%
\includegraphics{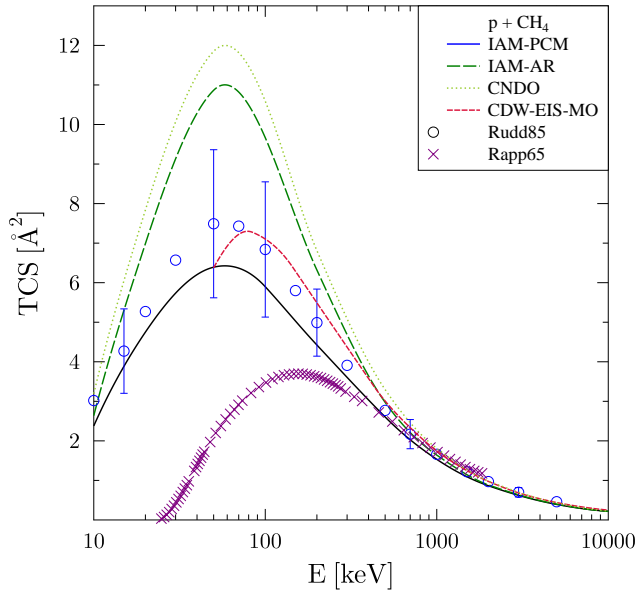}
}
\caption{%
Total cross section for net ionization in p-CH$_4$ collisions as a function
of impact energy.
IAM-PCM, IAM-AR, and CNDO: present calculations, CDW-EIS-MO~\cite{gulyas13};
experiments: Rudd85~\cite{Rudd85a}, Rapp65~\cite{rapp65} for electron impact using
equivelocity conversion.
}
\label{fig:ch4-ion}
\end{center}
\end{figure}

Figure~\ref{fig:ch4-ion} shows that this is indeed the case. 
However, the enhancement is relatively small (approximately 10\%) and insignificant for the
comparison with the experimental data and the other calculations included in the figure.
At impact energies $E\ge 700$ keV all data, including the IAM-AR and CNDO results, are in
reasonable agreement, corroborating the previous conclusion that
molecular structure effects are unimportant in this region~\cite{galassi00}. Toward lower energies the present IAM-AR and CNDO
results overestimate the data recommended by Rudd and coworkers~\cite{Rudd85a}, except at very low
energies in the $E=10-20$ keV range where
ionization is a relatively weak process. This implies that the atomic cross section overlaps
are small in this region, and indeed, IAM-AR and IAM-PCM results appear to merge, similarly to what is observed
at high energies ($E > 1$ MeV).

\begin{table}
\caption{
\label{tab:ch4}
Atomic occupation numbers $n_i^{\rm ao}$, gross atomic populations $\xi_i$, and atomic binding energies 
$\varepsilon_i$ (in a.u.) for CH$_4$
($i=1,\ldots ,4$)
The gross populations are obtained from the information provided in Table~III of~\cite{galassi00} 
whch in turn is based on the Mulliken population analysis of~\cite{hoffmann63}. 
The binding energies for atomic carbon 
are from the optimized effective potential calculations of~\cite{aashamer78} on which the present TC-BGM
cross section calculations are based.
}
\begin{tabular}{ccccc}
\hline \hline
 & H($1s$) & C($1s$) & C($2s$) & C($2p$)  \\
\hline
 $n_i^{\rm ao}$ & 4 & 2 & 2 & 2  \\
 $\xi_i$ & 3.468 & 2.0 & 1.133 & 3.399  \\
 $\varepsilon_i$ & 0.5 & 10.353 & 0.750 & 0.431 \\
\hline \hline
\end{tabular}
\end{table}

At intermediate energies the atomic cross sections are sufficiently large for
overlap effects to be substantial. Around the maximum the IAM-PCM net ionization cross section is 
less than 60\% of the IAM-AR value and in agreement with the recommended data 
within the reported uncertainties. 
A continuum distorted-wave eikonal initial-state (CDW-EIS) calculation in which 
instead of the IAM or the CNDO approach a spherical-basis representation of the
initial-state MOs was used~\cite{gulyas13}  
also agrees with the
data, but results in somewhat larger cross section values than the IAM-PCM and shows a different energy
dependence
with the maximum shifted toward higher energies. The latter might at least in part
be due to the limited validity of the perturbative CDW-EIS model at impact energies $E < 100$ keV
where electron capture gains importance as a competing reaction channel.

We have also included (absolute) electron impact measurements from~\cite{rapp65} in Fig.~\ref{fig:ch4-ion}
using equivelocity conversion. These measurements agree very well with the recommended data for proton impact 
above $E=500$ keV, i.e., at projectile speeds $v\ge 4.47$ a.u. or charge-magnitude-to-speed
ratios $\eta = 1/v \le 0.22$ a.u. One can infer from this
agreement that first-order perturbation theory in which cross sections do not depend on the sign of
the projectile charge is valid in this region. 
Toward lower energies, higher-order contributions become important, as do projectile mass and, for
electron projectiles, exchange effects, all of which contribute to the
differences in the electron- vs. proton-impact
cross sections.
It is interesting to
see how the CDW-EIS-MO calculation of Ref.~\cite{gulyas13} appears to be able to capture the
higher-order contributions quite well and agrees with the (proton-impact) measurements
down to energies which correspond
to $\eta$-values between 0.5 and 0.6.

\begin{figure}
\begin{center}
\resizebox{0.6\textwidth}{!}{%
\includegraphics{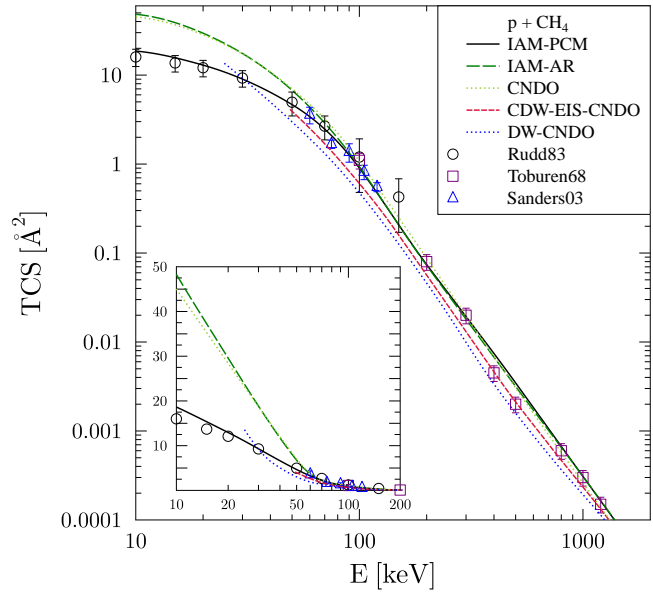}
}
\caption{%
Total cross section for net capture in p-CH$_4$ collisions as a function
of impact energy.
IAM-PCM, IAM-AR, and CNDO: present calculations, CDW-EIS-CNDO~\cite{quinto18}, DW-CNDO~\cite{purkait19};
experiments: Rudd83~\cite{Rudd83}, Toburen68~\cite{Toburen68}, Sanders03:~\cite{sanders03}.
}
\label{fig:ch4-cap}
\end{center}
\end{figure}

Figure~\ref{fig:ch4-cap} shows results for net capture using both double- and  
single-logarithmic scales to emphasize different impact energy regions. The present IAM-AR and CNDO
calculations yield very similar results at all energies and agree nicely with the experimental
data at $E\ge 50$ keV.
The close proximity of both calculations indicates that, unlike ionization,
capture does not simply increase with decreasing binding energy and more than one initially
occupied AO may contribute significantly. The details of the calculations show that at low impact
energies the resonant capture from H($1s$) dominates, while capture from
the carbon AOs gains relative importance toward higher energies. The differences in the orbital-specific
cross sections and those between the gross populations and occupation numbers listed in Table~\ref{tab:ch4} 
balance out and result in very similar IAM-AR and CNDO cross sections.

Figure~\ref{fig:ch4-cap} also includes two previous CNDO calculations; the calculation by Quinto and
co-workers~\cite{quinto18} uses the CDW-EIS model and the one by Purkait {\it et al.} \cite{purkait19}
a different
distorted-wave (DW) approach to calculate the AO-specific cross section contributions. Both works yield
somewhat smaller cross section values than the present CNDO calculation, but
show a similar impact energy dependence, at least
down to $E=50$ keV where the CDW-EIS calculation terminates. The cross-section curve from~\cite{purkait19}
extends further down to $E=25$ keV and indicates an even stronger increase
toward low energies than the present CNDO calculation, which significantly overestimates
the measurements in this region. We note that Purkait {\it et al.} also reported IAM-AR
calculations in their work which for p-CH$_4$ collisions are in close agreement with their
CNDO results.
Similarly, an earlier publication of the Argentinian group reported very similar
CDW-EIS IAM-AR and CNDO capture cross sections for a number of molecules, including CH$_4$~\cite{galassi10},
confirming our present finding regarding the comparison of both descriptions.
This suggests that the choice of {\it molecular} binding energies in the construction of the final
states in the perturbative models has but a very small influence on the magnitude of the
total capture cross section.

The present IAM-PCM results agree with the IAM-AR and CNDO calculations at high energies.
They begin to deviate from the latter
at $E \approx 70$ keV and follow 
the experimental data closely down to $E=10$ keV where the IAM-AR capture cross section is more than
twice as large. This is similar to what was found for other collision systems such as
p-CO and p-H$_2$O~\cite{hjl16} and gives us confidence in the validity of the IAM-PCM.
The key to its success from low- to high-velocity collisions is the impact-energy dependence 
of the weight factors in Eq.~(\ref{eq:pcm3}).
By contrast, the applicability of methods with energy-{\it independent} weights 
such as the IAM-AR and the CNDO approach
appears to be more limited.

We now turn to the DNA/RNA nucleobases
adenine, cytosine, guanine, thymine, and uracil.
Since the observations and conclusions are very similar for all five targets, we restrict the graphical discussion and
comparison with previous results to the 
proton-adenine collision system and provide our IAM-PCM results for the other target molecules
in tabular form only. The molecular geometry information required for these calculations is
taken from data available through the Molview project~\cite{molview}.

Experimental data for proton impact are rather scarce. For this reason, we include the
recent electron-impact ionization measurements of Ref.~\cite{rahman16} in the discussion.
We do not compare the present results with the (proton-impact) cross sections obtained from
a previous model calculation which combines the classical-trajectory Monte Carlo (CTMC) method with the classical
over barrier model (COB) ~\cite{Lekadir09} to avoid
overburdening the figures. The reader is referred to Refs.~\cite{Galassi12, Champion12} for comparisons
of the CTMC-COB results with perturbative CNDO
calculations and the measurements for proton impact.

\begin{figure}
\begin{center}
\resizebox{0.6\textwidth}{!}{%
\includegraphics{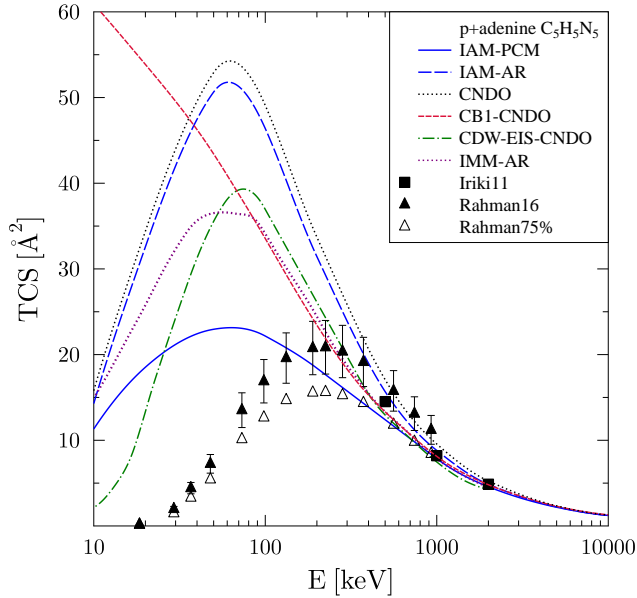}
}
\caption{%
Total cross section for net ionization in p-adenine (C$_5$H$_5$N$_5$) collisions as a function of impact energy.
IAM-PCM, IAM-AR, and CNDO: present calculations, CB1-CNDO and CDW-EIS-CNDO~\cite{Galassi12}, IMM-AR~\cite{paredes15};
experiments: Iriki11~\cite{iriki11a, iriki11b}, Rahman16~\cite{rahman16} for electron impact using equivelocity
conversion, Rahman75\%~are the data of~\cite{rahman16} multiplied by 0.75.
}
\label{fig:adenine-ion}
\end{center}
\end{figure}

Figure~\ref{fig:adenine-ion} shows the net ionization cross section for adenine target molecules.
We note that this cross section was measured by Tabet {\it et al.} at $E=80$ keV~\cite{Tabet10b},
but the data point is so high (at 155 \AA$^2$) that it is outside the scale of the figure.
The only other measurements for proton impact were reported by Iriki {\it et al.} \cite{iriki11a, iriki11b}.
Their data are included in Fig.~\ref{fig:adenine-ion} and appear to be 
somewhat lower than the electron-impact
measurements of \cite{rahman16}. If we multiply the latter by a factor of 0.75 they almost perfectly
match the proton data point at $E=1$ MeV, which is well described by most of the theoretical
calculations included in the figure. Toward lower $E$, the renormalized electron-impact data
agree very well with the present IAM-PCM calculations in the energy range in which one would
expect the proton- and electron-impact ionization cross section to be indistinguishable or very
nearly so.

By contrast, all other calculations predict a larger cross section below $E\approx 1$ MeV.
Similarly to the proton-methane case the present IAM-AR
results are slightly below the CNDO and overestimate the IAM-PCM cross section values
significantly (by more than a factor of two around the maximum). 
Some indirect confirmation that the overlaps incorporated into the IAM-PCM mimic
molecular effects appropriately comes from the independent-molecule-model (IMM) AR results included in
Fig.~\ref{fig:adenine-ion}. In this model, experimental cross sections for small
molecules are used to assemble the cross section for the larger molecule of interest~\cite{paredes15}. 
The results for adenine lie in between the IAM-PCM and IAM-AR cross sections suggesting that one can view
the IMM-AR as capturing overlap effects partially, i.e., within the small molecules used to assemble
the adenine ionization cross section, while overlaps between those small-molecule cross sections
are not accounted for.  

The present CNDO calculations are based on the Mulliken populations provided in Ref.~\cite{Galassi12},
which were also used for the CDW-EIS and the first-Born with corrected boundary conditions (CB1)
calculations reported in the same paper and included in Fig.~\ref{fig:adenine-ion}. 
The two perturbative calculations are in reasonable 
agreement with each other down to $E\approx 70$ keV where the CDW-EIS cross section assumes
its maximum, while the CB1 cross section keeps increasing toward lower energies. Both methods
predict significantly smaller cross section values 
than the present CNDO calculations in most of the impact energy interval shown in 
Fig.~\ref{fig:adenine-ion}. This may be due to (i) the perturbative frameworks used 
in the CDW-EIS and CB1 methods vs. the nonperturbative nature of the TC-BGM, 
(ii) the fact that in the perturbative
CNDO calculations the AO-specific cross sections include some molecular information 
through the choice of the final states, while the present
calculations do not. Given the analysis of net capture in p-CH$_4$ collisions presented above
the latter is unlikely,
but it would be interesting to see a CDW-EIS or CB1
calculation which uses atomic instead of molecular energies 
for the determination of the final (continuum) states to settle this issue.

\begin{figure}
\begin{center}
\resizebox{0.6\textwidth}{!}{%
\includegraphics{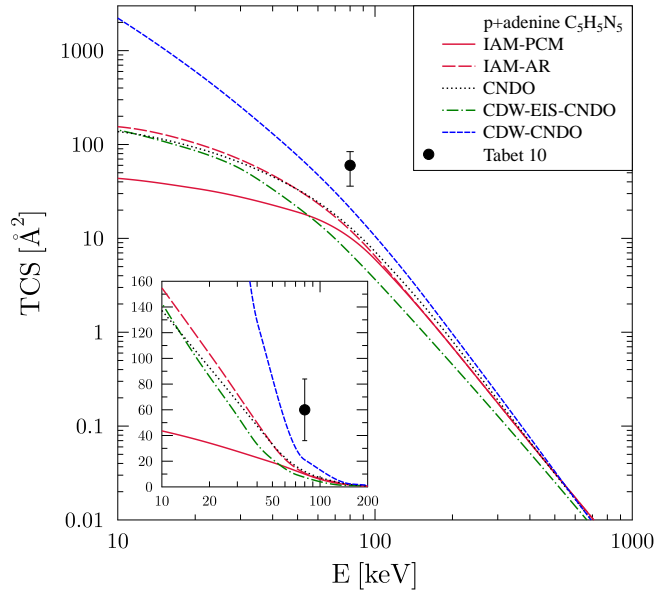}
}
\caption{%
Total cross section for net capture in p-adenine (C$_5$H$_5$N$_5$) collisions as a function of impact energy.
IAM-PCM, IAM-AR, and CNDO: present calculations, CDW-EIS-CNDO and CDW-CNDO~\cite{Champion12};
experiment: Tabet10~\cite{Tabet10b}.
}
\label{fig:adenine-cap}
\end{center}
\end{figure}

Results for net capture are shown in Fig.~\ref{fig:adenine-cap}.
We compare the present CNDO, IAM-AR and IAM-PCM data with the
CDW and CDW-EIS CNDO calculations of Ref.~\cite{Champion12} and the only
reported experimental data point at $E=80$ keV~\cite{Tabet10b}. 
The CDW-EIS model produces lower cross section values than
the present CNDO calculation at all impact energies except around $E=10$ keV which is probably outside the
region of validity of the perturbative model.
By contrast, the CDW model, which differs from the CDW-EIS in the choice of the distortion
factor in the initial
state, results in much larger cross section values which merge with the other
theoretical results only at energies $E\ge 500$ keV.

As for p-CH$_4$ collisions, the IAM-AR
is in good agreement with the present CNDO results over most of the energy range shown. Only below
$E\approx 40$ keV do the results of both models deviate somewhat more strongly than for methane.
This is not surprising given
the larger number of contributing electron subshells in adenine, which makes it less likely that the
differences in orbital-specific capture cross sections and gross populations vs. atomic occupation
numbers balance out.

The experimental data point of~\cite{Tabet10b} at $E=80$ keV is higher than all theoretical
results included in Fig.~\ref{fig:adenine-cap}. The discrepancy with the CDW CNDO calculation
is perhaps acceptable, but this may be fortuitous given that the CDW model is sometimes
considered inferior to the CDW-EIS because of its use of non-normalized initial-state wave 
functions~\cite{gulyas12}.
The authors of Ref.~\cite{Champion12} deemed the seemingly good performance of the
CDW model ``unexpected'' and concluded that new measurements would be welcome.
The present calculations reinforce the latter point.
In particular, it
would be of great interest if new measurements were extended to lower impact energies
where the IAM-PCM predicts a much smaller capture cross section
than all other theoretical models. 
Given the good agreement of the IAM-PCM net capture 
with experimental data for smaller molecules such as CH$_4$ (cf. Fig.~\ref{fig:ch4-cap})
it would be surprising if the prediction for proton-adenine collisions
would be off.

We have checked that the situation is similar for the other DNA/RNA nucleobases.
In fact, all the features seen in Figs.~\ref{fig:adenine-ion} and~\ref{fig:adenine-cap}
for adenine are also present for cytosine, guanine, thymine and uracil. 
This can be explained by the similar atomic building blocks and structures of these molecules.
In lieu of figures we provide tables with the present
IAM-PCM results: Table~\ref{tab:Pyr}  lists the net capture and ionization cross
sections for the pyrimidines cytosine, thymine, and uracil and Table~\ref{tab:Pur} 
the results for the purines adenine and guanine. 
We note that the ionization cross sections were previously included in tables
presented in~\cite{hjl19}, in which scaling properties were studied for
larger classes of systems. We repeat these data here for the convenience of the reader.

\begin{table}[h]
 \caption{Orientation-averaged IAM-PCM net capture and ionization cross sections for proton collisions with 
 the pyrimidines cytosine, thymine, and uracil (in \AA$^2$).}
 \label{tab:Pyr}
 \begin{tabular}{rcccccc}
  \hline \hline
  	&	 \multicolumn{2}{c} {Cytosine  (C$_4$H$_5$N$_3$O)} &    
		  \multicolumn{2}{c} {Thymine  (C$_5$H$_6$N$_2$O$_2$)} & 
		   \multicolumn{2}{c} { Uracil (C$_4$H$_4$N$_2$O$_2$) }  \\
  E [keV]    & Capture		& Ionization	& Capture		& Ionization	& Capture		& Ionization	\\         
  \hline
   10 	  	  & 37.71	& 9.34	 & 41.03	& 10.62	 & 35.80	& 8.92 \\	
   20 	 	  & 29.32	& 15.42	 &32.28 	& 17.41	 & 27.74	& 14.83 \\	
   50 	 	  & 15.95	& 19.44	 & 18.00	& 21.93	 & 15.44	& 18.90 \\	
  100 	  & 4.99	& 18.75	 & 5.58	& 21.17	 & 4.94	& 18.23 \\	
  200 	  &0.61 	& 15.49	 & 0.69	& 17.63	 &0.65 	& 15.13 \\	
  500 	  & 0.029	& 10.28	 &0.031 	& 11.72	 &0.028 	& 9.99 \\	
 1000 	  & 0.0026	& 6.46	 & 0.0029	& 7.35	 &0.0026 	& 6.21 \\	
 2000 	  & 		& 3.76	 & 		& 4.26	 & 		& 3.65 \\	
 5000 	  & 		& 1.80	 & 		& 2.03	 & 		& 1.76 \\	
10000 	  & 		& 1.02	 & 		& 1.14	 &		& 0.98 \\              
  \hline \hline
 \end{tabular}
\end{table}
\begin{table}[h]
 \caption{Orientation-averaged IAM-PCM net capture and ionization cross sections for proton collisions with 
  the purines adenine and guanine (in \AA$^2$).}
 \label{tab:Pur}
 \begin{tabular}{rcccc}
  \hline \hline
  		 &   \multicolumn{2}{c} {Adenine  (C$_5$H$_5$N$_5$)} & 
		       \multicolumn{2}{c} { Guanine  (C$_5$H$_5$N$_5$O)}     \\
  E [keV]    & Capture		& Ionization	      & Capture		& Ionization\\
  \hline
   10 	 	 & 43.59	  & 11.32 &45.12	   & 12.04  \\	
   20 	  	 & 34.12	  & 18.53 &35.58	   & 19.70  \\	
   50 	  	 & 19.04	  & 22.96 &20.50	   & 24.52  \\	
  100 	 & 5.98	  & 22.17 &6.79	   	   & 23.68  \\	
  200 	 & 0.69	  & 18.41 &0.82	            & 19.91  \\	
  500 	 & 0.032	  & 12.40 &0.036	   & 13.33  \\	
 1000 	 & 0.0033	  & 7.86   &0.0036	   & 8.47    \\	
 2000 	 & 		  & 4.58   &	   	   & 4.96    \\	
 5000 	 & 		  & 2.18   &	  	   & 2.39    \\	
10000 	 & 		  & 1.23   &	  	   & 1.33    \\              
  \hline \hline
 \end{tabular}
\end{table}

\section{Conclusions}
\label{sec:conclusions}
We have presented IAM-PCM calculations for 
proton collisions from methane molecules and the nucleobases adenine, cytosine, guanine, thymine, and uracil over wide ranges
of impact energy from $E=10$ keV to $E=1$ MeV for net capture and up to $E=10$ MeV for net ionization.
Like the simpler IAM-AR and the widely-used CNDO approach the IAM-PCM is based on atomic cross-section calculations,
but in contrast to the former it weighs the atomic contributions in an impact-energy-dependent way.
The weight factors are obtained from the geometrical overlaps which arise when
one pictures the atomic cross sections as circular disks in the impact-parameter plane. 
The overlaps can be substantial thereby leading to a significant reduction of
IAM-PCM compared to IAM-AR cross sections. This effect is most pronounced for electron capture at low energies where
we found
discrepancies between IAM-PCM and IAM-AR cross sections of
up to a factor of three to four. We have shown these discrepancies for proton-adenine 
collisions only, but have checked that they are similar for the other nucleobases.
New measurements are required to test these predictions.

In the case of ionization the discrepancies between different
theoretical models are less dramatic, but they are sizable, especially around
the cross section maximum. Again, these trends are similar for all nucleobases and an experimental
study that would test this similarity in a systematic way would be of great interest.
As for capture, the only existing experimental data points
in this region appear to be too high. 
 
In contrast to the IAM-PCM, the CNDO approach, when coupled with the present (nonperturbative)
atomic-orbital-specific cross section calculations, does not lead to significant differences
to the simple IAM-AR.
Previous (perturbative) work provided some evidence that the differences between both models
become more pronounced when differential cross sections are calculated~\cite{galassi00}.
In the context of the present analysis it would be interesting to know if the deviations in
the differential cross-section results are
related to the use of molecular instead of atomic binding energies in the determination of the final continuum
states in the perturbative CNDO calculations. On the level of total cross sections this choice
appears to be of minor importance.

Our own future work will focus on IAM-PCM studies of collisions involving multiply-charged ions. 
Preliminary calculations show, not surprisingly, that the overlap effect is stronger and IAM-PCM
total cross sections merge with IAM-AR results at higher impact energies than for proton impact.
The role of multi-electron processes will be enhanced as well. These processes require 
an extension of the model to allow for the calculation of impact-parameter-dependent probabilities
which can be fed into a multinomial analysis of multiple capture and ionization processes.
Work in this direction is in progress. 

\begin{acknowledgments}
This work was supported by the Natural Sciences and Engineering Research Council of Canada (NSERC).
One of us (H. J. L.) would like to thank the Center for Scientific Computing, University of Frankfurt 
for making their
High Performance Computing facilities available.
\end{acknowledgments}

%
% BibTeX users please use
%\bibliographystyle{plain}
%\bibliographystyle{unsrt}
\bibliography{pcm-2019}

\begin{thebibliography}{32}%
\makeatletter
\providecommand \@ifxundefined [1]{%
 \@ifx{#1\undefined}
}%
\providecommand \@ifnum [1]{%
 \ifnum #1\expandafter \@firstoftwo
 \else \expandafter \@secondoftwo
 \fi
}%
\providecommand \@ifx [1]{%
 \ifx #1\expandafter \@firstoftwo
 \else \expandafter \@secondoftwo
 \fi
}%
\providecommand \natexlab [1]{#1}%
\providecommand \enquote  [1]{``#1''}%
\providecommand \bibnamefont  [1]{#1}%
\providecommand \bibfnamefont [1]{#1}%
\providecommand \citenamefont [1]{#1}%
\providecommand \href@noop [0]{\@secondoftwo}%
\providecommand \href [0]{\begingroup \@sanitize@url \@href}%
\providecommand \@href[1]{\@@startlink{#1}\@@href}%
\providecommand \@@href[1]{\endgroup#1\@@endlink}%
\providecommand \@sanitize@url [0]{\catcode `\\12\catcode `\$12\catcode
  `\&12\catcode `\#12\catcode `\^12\catcode `\_12\catcode `\%12\relax}%
\providecommand \@@startlink[1]{}%
\providecommand \@@endlink[0]{}%
\providecommand \url  [0]{\begingroup\@sanitize@url \@url }%
\providecommand \@url [1]{\endgroup\@href {#1}{\urlprefix }}%
\providecommand \urlprefix  [0]{URL }%
\providecommand \Eprint [0]{\href }%
\providecommand \doibase [0]{http://dx.doi.org/}%
\providecommand \selectlanguage [0]{\@gobble}%
\providecommand \bibinfo  [0]{\@secondoftwo}%
\providecommand \bibfield  [0]{\@secondoftwo}%
\providecommand \translation [1]{[#1]}%
\providecommand \BibitemOpen [0]{}%
\providecommand \bibitemStop [0]{}%
\providecommand \bibitemNoStop [0]{.\EOS\space}%
\providecommand \EOS [0]{\spacefactor3000\relax}%
\providecommand \BibitemShut  [1]{\csname bibitem#1\endcsname}%
\let\auto@bib@innerbib\@empty
%</preamble>
\bibitem [{\citenamefont {Bragg}\ and\ \citenamefont
  {Kleeman}(1905)}]{bragg05}%
  \BibitemOpen
  \bibfield  {author} {\bibinfo {author} {\bibfnamefont {W.~H.}\ \bibnamefont
  {Bragg}}\ and\ \bibinfo {author} {\bibfnamefont {R.}~\bibnamefont
  {Kleeman}},\ }\href {https://doi.org/10.1080/14786440509463378} {\bibfield
  {journal} {\bibinfo  {journal} {Phil. Mag.}\ }\textbf {\bibinfo {volume}
  {10}},\ \bibinfo {pages} {318} (\bibinfo {year} {1905})}\BibitemShut
  {NoStop}%
\bibitem [{\citenamefont {Otvos}\ and\ \citenamefont
  {Stevenson}(1956)}]{otvos56}%
  \BibitemOpen
  \bibfield  {author} {\bibinfo {author} {\bibfnamefont {J.~W.}\ \bibnamefont
  {Otvos}}\ and\ \bibinfo {author} {\bibfnamefont {D.~P.}\ \bibnamefont
  {Stevenson}},\ }\href {https://doi.org/10.1021/ja01584a009} {\bibfield
  {journal} {\bibinfo  {journal} {J. Am. Chem. Soc.}\ }\textbf {\bibinfo
  {volume} {78}},\ \bibinfo {pages} {546} (\bibinfo {year} {1956})}\BibitemShut
  {NoStop}%
\bibitem [{\citenamefont {Galassi}\ \emph {et~al.}(2000)\citenamefont
  {Galassi}, \citenamefont {Rivarola}, \citenamefont {Beuve}, \citenamefont
  {Olivera},\ and\ \citenamefont {Fainstein}}]{galassi00}%
  \BibitemOpen
  \bibfield  {author} {\bibinfo {author} {\bibfnamefont {M.~E.}\ \bibnamefont
  {Galassi}}, \bibinfo {author} {\bibfnamefont {R.~D.}\ \bibnamefont
  {Rivarola}}, \bibinfo {author} {\bibfnamefont {M.}~\bibnamefont {Beuve}},
  \bibinfo {author} {\bibfnamefont {G.~H.}\ \bibnamefont {Olivera}}, \ and\
  \bibinfo {author} {\bibfnamefont {P.~D.}\ \bibnamefont {Fainstein}},\ }\href
  {\doibase 10.1103/PhysRevA.62.022701} {\bibfield  {journal} {\bibinfo
  {journal} {Phys. Rev. A}\ }\textbf {\bibinfo {volume} {62}},\ \bibinfo
  {pages} {022701} (\bibinfo {year} {2000})}\BibitemShut {NoStop}%
\bibitem [{\citenamefont {Blanco}\ and\ \citenamefont
  {Garc\'{i}a}(2003)}]{Blanco2003}%
  \BibitemOpen
  \bibfield  {author} {\bibinfo {author} {\bibfnamefont {F.}~\bibnamefont
  {Blanco}}\ and\ \bibinfo {author} {\bibfnamefont {G.}~\bibnamefont
  {Garc\'{i}a}},\ }\href {https://doi.org/10.1016/j.physleta.2003.09.016}
  {\bibfield  {journal} {\bibinfo  {journal} {Phys. Lett. A}\ }\textbf
  {\bibinfo {volume} {317}},\ \bibinfo {pages} {458} (\bibinfo {year}
  {2003})}\BibitemShut {NoStop}%
\bibitem [{\citenamefont {Deutsch}\ \emph {et~al.}(2000)\citenamefont
  {Deutsch}, \citenamefont {Becker}, \citenamefont {Matt},\ and\ \citenamefont
  {M\"ark}}]{Deutsch00}%
  \BibitemOpen
  \bibfield  {author} {\bibinfo {author} {\bibfnamefont {H.}~\bibnamefont
  {Deutsch}}, \bibinfo {author} {\bibfnamefont {K.}~\bibnamefont {Becker}},
  \bibinfo {author} {\bibfnamefont {S.}~\bibnamefont {Matt}}, \ and\ \bibinfo
  {author} {\bibfnamefont {T.~D.}\ \bibnamefont {M\"ark}},\ }\href
  {https://doi.org/10.1016/S1387-3806(99)00257-2} {\bibfield  {journal}
  {\bibinfo  {journal} {Int. J. Mass Spectrom.}\ }\textbf {\bibinfo {volume}
  {197}},\ \bibinfo {pages} {37} (\bibinfo {year} {2000})}\BibitemShut
  {NoStop}%
\bibitem [{\citenamefont {Champion}\ \emph {et~al.}(2012)\citenamefont
  {Champion}, \citenamefont {Weck}, \citenamefont {Lekadir}, \citenamefont
  {Galassi}, \citenamefont {Foj\'{o}n}, \citenamefont {Abufager}, \citenamefont
  {Rivarola},\ and\ \citenamefont {Hanssen}}]{Champion12}%
  \BibitemOpen
  \bibfield  {author} {\bibinfo {author} {\bibfnamefont {C.}~\bibnamefont
  {Champion}}, \bibinfo {author} {\bibfnamefont {P.~F.}\ \bibnamefont {Weck}},
  \bibinfo {author} {\bibfnamefont {H.}~\bibnamefont {Lekadir}}, \bibinfo
  {author} {\bibfnamefont {M.~E.}\ \bibnamefont {Galassi}}, \bibinfo {author}
  {\bibfnamefont {O.~A.}\ \bibnamefont {Foj\'{o}n}}, \bibinfo {author}
  {\bibfnamefont {P.}~\bibnamefont {Abufager}}, \bibinfo {author}
  {\bibfnamefont {R.~D.}\ \bibnamefont {Rivarola}}, \ and\ \bibinfo {author}
  {\bibfnamefont {J.}~\bibnamefont {Hanssen}},\ }\href {\doibase
  10.1088/0031-9155/57/10/3039} {\bibfield  {journal} {\bibinfo  {journal}
  {Phys. Med. Biol.}\ }\textbf {\bibinfo {volume} {57}},\ \bibinfo {pages}
  {3039} (\bibinfo {year} {2012})}\BibitemShut {NoStop}%
\bibitem [{\citenamefont {Quinto}\ \emph {et~al.}(2018)\citenamefont {Quinto},
  \citenamefont {Montenegro}, \citenamefont {Monti}, \citenamefont
  {Foj\'{o}n},\ and\ \citenamefont {Rivarola}}]{quinto18}%
  \BibitemOpen
  \bibfield  {author} {\bibinfo {author} {\bibfnamefont {M.~A.}\ \bibnamefont
  {Quinto}}, \bibinfo {author} {\bibfnamefont {P.~R.}\ \bibnamefont
  {Montenegro}}, \bibinfo {author} {\bibfnamefont {J.~M.}\ \bibnamefont
  {Monti}}, \bibinfo {author} {\bibfnamefont {O.~A.}\ \bibnamefont
  {Foj\'{o}n}}, \ and\ \bibinfo {author} {\bibfnamefont {R.~D.}\ \bibnamefont
  {Rivarola}},\ }\href {https://doi.org/10.1088/1361-6455/aad152} {\bibfield
  {journal} {\bibinfo  {journal} {J. Phys. B}\ }\textbf {\bibinfo {volume}
  {51}},\ \bibinfo {pages} {165201} (\bibinfo {year} {2018})}\BibitemShut
  {NoStop}%
\bibitem [{\citenamefont {Mulliken}(1955)}]{mulliken55}%
  \BibitemOpen
  \bibfield  {author} {\bibinfo {author} {\bibfnamefont {R.~S.}\ \bibnamefont
  {Mulliken}},\ }\href {https://doi.org/10.1063/1.1740588} {\bibfield
  {journal} {\bibinfo  {journal} {J. Chem. Phys.}\ }\textbf {\bibinfo {volume}
  {23}},\ \bibinfo {pages} {1833} (\bibinfo {year} {1955})}\BibitemShut
  {NoStop}%
\bibitem [{\citenamefont {L\"udde}\ \emph {et~al.}(2016)\citenamefont
  {L\"udde}, \citenamefont {Achenbach}, \citenamefont {Kalkbrenner},
  \citenamefont {Jankowiak},\ and\ \citenamefont {Kirchner}}]{hjl16}%
  \BibitemOpen
  \bibfield  {author} {\bibinfo {author} {\bibfnamefont {H.~J.}\ \bibnamefont
  {L\"udde}}, \bibinfo {author} {\bibfnamefont {A.}~\bibnamefont {Achenbach}},
  \bibinfo {author} {\bibfnamefont {T.}~\bibnamefont {Kalkbrenner}}, \bibinfo
  {author} {\bibfnamefont {H.-C.}\ \bibnamefont {Jankowiak}}, \ and\ \bibinfo
  {author} {\bibfnamefont {T.}~\bibnamefont {Kirchner}},\ }\href
  {https://doi.org/10.1140/epjd/e2016-70097-5} {\bibfield  {journal} {\bibinfo
  {journal} {Eur. Phys. J. D}\ }\textbf {\bibinfo {volume} {70}},\ \bibinfo
  {pages} {82} (\bibinfo {year} {2016})}\BibitemShut {NoStop}%
\bibitem [{\citenamefont {L\"udde}\ \emph {et~al.}(2018)\citenamefont
  {L\"udde}, \citenamefont {Horbatsch},\ and\ \citenamefont
  {Kirchner}}]{hjl18}%
  \BibitemOpen
  \bibfield  {author} {\bibinfo {author} {\bibfnamefont {H.~J.}\ \bibnamefont
  {L\"udde}}, \bibinfo {author} {\bibfnamefont {M.}~\bibnamefont {Horbatsch}},
  \ and\ \bibinfo {author} {\bibfnamefont {T.}~\bibnamefont {Kirchner}},\
  }\href {https://doi.org/10.1140/epjb/e2018-90165-x} {\bibfield  {journal}
  {\bibinfo  {journal} {Eur. Phys. J. B}\ }\textbf {\bibinfo {volume} {91}},\
  \bibinfo {pages} {99} (\bibinfo {year} {2018})}\BibitemShut {NoStop}%
\bibitem [{\citenamefont {Kirchner}\ \emph {et~al.}(1998)\citenamefont
  {Kirchner}, \citenamefont {Guly\'as}, \citenamefont {L\"udde}, \citenamefont
  {Engel},\ and\ \citenamefont {Dreizler}}]{tom98}%
  \BibitemOpen
  \bibfield  {author} {\bibinfo {author} {\bibfnamefont {T.}~\bibnamefont
  {Kirchner}}, \bibinfo {author} {\bibfnamefont {L.}~\bibnamefont {Guly\'as}},
  \bibinfo {author} {\bibfnamefont {H.~J.}\ \bibnamefont {L\"udde}}, \bibinfo
  {author} {\bibfnamefont {E.}~\bibnamefont {Engel}}, \ and\ \bibinfo {author}
  {\bibfnamefont {R.~M.}\ \bibnamefont {Dreizler}},\ }\href {\doibase
  10.1103/PhysRevA.58.2063} {\bibfield  {journal} {\bibinfo  {journal} {Phys.
  Rev. A}\ }\textbf {\bibinfo {volume} {58}},\ \bibinfo {pages} {2063}
  (\bibinfo {year} {1998})}\BibitemShut {NoStop}%
\bibitem [{\citenamefont {L\"udde}\ \emph {et~al.}(2019)\citenamefont
  {L\"udde}, \citenamefont {Horbatsch},\ and\ \citenamefont
  {Kirchner}}]{hjl19}%
  \BibitemOpen
  \bibfield  {author} {\bibinfo {author} {\bibfnamefont {H.~J.}\ \bibnamefont
  {L\"udde}}, \bibinfo {author} {\bibfnamefont {M.}~\bibnamefont {Horbatsch}},
  \ and\ \bibinfo {author} {\bibfnamefont {T.}~\bibnamefont {Kirchner}},\
  }\href@noop {} {\  (\bibinfo {year} {2019})},\ \bibinfo {note} {submitted to
  J. Phys. B.},\ \Eprint {http://arxiv.org/abs/1905.02273} {arXiv:1905.02273
  [physics.atm-clus]} \BibitemShut {NoStop}%
\bibitem [{\citenamefont {Zapukhlyak}\ \emph {et~al.}(2005)\citenamefont
  {Zapukhlyak}, \citenamefont {Kirchner}, \citenamefont {L\"udde},
  \citenamefont {Knoop}, \citenamefont {Morgenstern},\ and\ \citenamefont
  {Hoekstra}}]{tcbgm}%
  \BibitemOpen
  \bibfield  {author} {\bibinfo {author} {\bibfnamefont {M.}~\bibnamefont
  {Zapukhlyak}}, \bibinfo {author} {\bibfnamefont {T.}~\bibnamefont
  {Kirchner}}, \bibinfo {author} {\bibfnamefont {H.~J.}\ \bibnamefont
  {L\"udde}}, \bibinfo {author} {\bibfnamefont {S.}~\bibnamefont {Knoop}},
  \bibinfo {author} {\bibfnamefont {R.}~\bibnamefont {Morgenstern}}, \ and\
  \bibinfo {author} {\bibfnamefont {R.}~\bibnamefont {Hoekstra}},\ }\href
  {\doibase 10.1088/0953-4075/38/14/003} {\bibfield  {journal} {\bibinfo
  {journal} {J. Phys. B}\ }\textbf {\bibinfo {volume} {38}},\ \bibinfo {pages}
  {2353} (\bibinfo {year} {2005})}\BibitemShut {NoStop}%
\bibitem [{\citenamefont {Galassi}\ \emph {et~al.}(2010)\citenamefont
  {Galassi}, \citenamefont {Abufager}, \citenamefont {Fainstein},\ and\
  \citenamefont {Rivarola}}]{galassi10}%
  \BibitemOpen
  \bibfield  {author} {\bibinfo {author} {\bibfnamefont {M.~E.}\ \bibnamefont
  {Galassi}}, \bibinfo {author} {\bibfnamefont {P.~N.}\ \bibnamefont
  {Abufager}}, \bibinfo {author} {\bibfnamefont {P.~D.}\ \bibnamefont
  {Fainstein}}, \ and\ \bibinfo {author} {\bibfnamefont {R.~D.}\ \bibnamefont
  {Rivarola}},\ }\href {\doibase 10.1103/PhysRevA.81.062713} {\bibfield
  {journal} {\bibinfo  {journal} {Phys. Rev. A}\ }\textbf {\bibinfo {volume}
  {81}},\ \bibinfo {pages} {062713} (\bibinfo {year} {2010})}\BibitemShut
  {NoStop}%
\bibitem [{\citenamefont {Hoffmann}(1963)}]{hoffmann63}%
  \BibitemOpen
  \bibfield  {author} {\bibinfo {author} {\bibfnamefont {R.}~\bibnamefont
  {Hoffmann}},\ }\href {https://doi.org/10.1063/1.1734456} {\bibfield
  {journal} {\bibinfo  {journal} {J. Chem. Phys.}\ }\textbf {\bibinfo {volume}
  {39}},\ \bibinfo {pages} {1397} (\bibinfo {year} {1963})}\BibitemShut
  {NoStop}%
\bibitem [{\citenamefont {Aashamer}\ \emph {et~al.}(1978)\citenamefont
  {Aashamer}, \citenamefont {Luke},\ and\ \citenamefont {Talman}}]{aashamer78}%
  \BibitemOpen
  \bibfield  {author} {\bibinfo {author} {\bibfnamefont {K.}~\bibnamefont
  {Aashamer}}, \bibinfo {author} {\bibfnamefont {T.~M.}\ \bibnamefont {Luke}},
  \ and\ \bibinfo {author} {\bibfnamefont {J.~D.}\ \bibnamefont {Talman}},\
  }\href {https://doi.org/10.1016/0092-640X(78)90019-0} {\bibfield  {journal}
  {\bibinfo  {journal} {At. Data Nucl. Data Tables}\ }\textbf {\bibinfo
  {volume} {22}},\ \bibinfo {pages} {443} (\bibinfo {year} {1978})}\BibitemShut
  {NoStop}%
\bibitem [{\citenamefont {Rudd}\ \emph {et~al.}(1985)\citenamefont {Rudd},
  \citenamefont {Kim}, \citenamefont {Madison},\ and\ \citenamefont
  {Gallagher}}]{Rudd85a}%
  \BibitemOpen
  \bibfield  {author} {\bibinfo {author} {\bibfnamefont {M.~E.}\ \bibnamefont
  {Rudd}}, \bibinfo {author} {\bibfnamefont {Y.~K.}\ \bibnamefont {Kim}},
  \bibinfo {author} {\bibfnamefont {D.~H.}\ \bibnamefont {Madison}}, \ and\
  \bibinfo {author} {\bibfnamefont {J.~W.}\ \bibnamefont {Gallagher}},\ }\href
  {\doibase 10.1103/RevModPhys.57.965} {\bibfield  {journal} {\bibinfo
  {journal} {Rev. Mod. Phys.}\ }\textbf {\bibinfo {volume} {57}},\ \bibinfo
  {pages} {965} (\bibinfo {year} {1985})}\BibitemShut {NoStop}%
\bibitem [{\citenamefont {Guly\'as}\ \emph {et~al.}(2013)\citenamefont
  {Guly\'as}, \citenamefont {T\'oth},\ and\ \citenamefont {Nagy}}]{gulyas13}%
  \BibitemOpen
  \bibfield  {author} {\bibinfo {author} {\bibfnamefont {L.}~\bibnamefont
  {Guly\'as}}, \bibinfo {author} {\bibfnamefont {I.}~\bibnamefont {T\'oth}}, \
  and\ \bibinfo {author} {\bibfnamefont {L.}~\bibnamefont {Nagy}},\ }\href
  {\doibase 10.1088/0953-4075/46/7/075201} {\bibfield  {journal} {\bibinfo
  {journal} {J. Phys. B}\ }\textbf {\bibinfo {volume} {46}},\ \bibinfo {pages}
  {075201} (\bibinfo {year} {2013})}\BibitemShut {NoStop}%
\bibitem [{\citenamefont {Rapp}\ and\ \citenamefont
  {Englander-Golden}(1965)}]{rapp65}%
  \BibitemOpen
  \bibfield  {author} {\bibinfo {author} {\bibfnamefont {D.}~\bibnamefont
  {Rapp}}\ and\ \bibinfo {author} {\bibfnamefont {P.}~\bibnamefont
  {Englander-Golden}},\ }\href {https://doi.org/10.1063/1.1696957} {\bibfield
  {journal} {\bibinfo  {journal} {J. Chem. Phys.}\ }\textbf {\bibinfo {volume}
  {43}},\ \bibinfo {pages} {1464} (\bibinfo {year} {1965})}\BibitemShut
  {NoStop}%
\bibitem [{\citenamefont {Purkait}\ \emph {et~al.}(2019)\citenamefont
  {Purkait}, \citenamefont {Samaddar}, \citenamefont {Halder}, \citenamefont
  {Mandal},\ and\ \citenamefont {Purkait}}]{purkait19}%
  \BibitemOpen
  \bibfield  {author} {\bibinfo {author} {\bibfnamefont {K.}~\bibnamefont
  {Purkait}}, \bibinfo {author} {\bibfnamefont {S.}~\bibnamefont {Samaddar}},
  \bibinfo {author} {\bibfnamefont {S.}~\bibnamefont {Halder}}, \bibinfo
  {author} {\bibfnamefont {C.~R.}\ \bibnamefont {Mandal}}, \ and\ \bibinfo
  {author} {\bibfnamefont {M.}~\bibnamefont {Purkait}},\ }\href
  {https://doi.org/10.1007/s13538-019-00669-2} {\bibfield  {journal} {\bibinfo
  {journal} {Braz. J. Phys.}\ }\textbf {\bibinfo {volume} {49}},\ \bibinfo
  {pages} {473} (\bibinfo {year} {2019})}\BibitemShut {NoStop}%
\bibitem [{\citenamefont {Rudd}\ \emph {et~al.}(1983)\citenamefont {Rudd},
  \citenamefont {DuBois}, \citenamefont {Toburen}, \citenamefont {Ratcliffe},\
  and\ \citenamefont {Goffe}}]{Rudd83}%
  \BibitemOpen
  \bibfield  {author} {\bibinfo {author} {\bibfnamefont {M.~E.}\ \bibnamefont
  {Rudd}}, \bibinfo {author} {\bibfnamefont {R.~D.}\ \bibnamefont {DuBois}},
  \bibinfo {author} {\bibfnamefont {L.~H.}\ \bibnamefont {Toburen}}, \bibinfo
  {author} {\bibfnamefont {C.~A.}\ \bibnamefont {Ratcliffe}}, \ and\ \bibinfo
  {author} {\bibfnamefont {T.~V.}\ \bibnamefont {Goffe}},\ }\href {\doibase
  10.1103/PhysRevA.28.3244} {\bibfield  {journal} {\bibinfo  {journal} {Phys.
  Rev. A}\ }\textbf {\bibinfo {volume} {28}},\ \bibinfo {pages} {3244}
  (\bibinfo {year} {1983})}\BibitemShut {NoStop}%
\bibitem [{\citenamefont {Toburen}\ \emph {et~al.}(1968)\citenamefont
  {Toburen}, \citenamefont {Nakai},\ and\ \citenamefont {Langley}}]{Toburen68}%
  \BibitemOpen
  \bibfield  {author} {\bibinfo {author} {\bibfnamefont {L.~H.}\ \bibnamefont
  {Toburen}}, \bibinfo {author} {\bibfnamefont {M.~Y.}\ \bibnamefont {Nakai}},
  \ and\ \bibinfo {author} {\bibfnamefont {R.~A.}\ \bibnamefont {Langley}},\
  }\href {\doibase 10.1103/PhysRev.171.114} {\bibfield  {journal} {\bibinfo
  {journal} {Phys. Rev.}\ }\textbf {\bibinfo {volume} {171}},\ \bibinfo {pages}
  {114} (\bibinfo {year} {1968})}\BibitemShut {NoStop}%
\bibitem [{\citenamefont {Sanders}\ \emph {et~al.}(2003)\citenamefont
  {Sanders}, \citenamefont {Varghese}, \citenamefont {Fleming},\ and\
  \citenamefont {Soosai}}]{sanders03}%
  \BibitemOpen
  \bibfield  {author} {\bibinfo {author} {\bibfnamefont {J.~M.}\ \bibnamefont
  {Sanders}}, \bibinfo {author} {\bibfnamefont {S.~L.}\ \bibnamefont
  {Varghese}}, \bibinfo {author} {\bibfnamefont {C.~H.}\ \bibnamefont
  {Fleming}}, \ and\ \bibinfo {author} {\bibfnamefont {G.~A.}\ \bibnamefont
  {Soosai}},\ }\href {\doibase 10.1088/0953-4075/36/18/311} {\bibfield
  {journal} {\bibinfo  {journal} {J. Phys. B}\ }\textbf {\bibinfo {volume}
  {36}},\ \bibinfo {pages} {3835} (\bibinfo {year} {2003})}\BibitemShut
  {NoStop}%
\bibitem [{mol()}]{molview}%
  \BibitemOpen
  \href@noop {} {}\bibinfo {note} {Mol{V}iew: http://molview.org [Online;
  accessed 2019-13-07]}\BibitemShut {NoStop}%
\bibitem [{\citenamefont {Rahman}\ and\ \citenamefont
  {Krishnakumar}(2016)}]{rahman16}%
  \BibitemOpen
  \bibfield  {author} {\bibinfo {author} {\bibfnamefont {M.~A.}\ \bibnamefont
  {Rahman}}\ and\ \bibinfo {author} {\bibfnamefont {E.}~\bibnamefont
  {Krishnakumar}},\ }\href {http://dx.doi.org/10.1063/1.4948412} {\bibfield
  {journal} {\bibinfo  {journal} {J. Chem. Phys.}\ }\textbf {\bibinfo {volume}
  {144}},\ \bibinfo {pages} {161102} (\bibinfo {year} {2016})}\BibitemShut
  {NoStop}%
\bibitem [{\citenamefont {Lekadir}\ \emph {et~al.}(2009)\citenamefont
  {Lekadir}, \citenamefont {Abbas}, \citenamefont {Champion}, \citenamefont
  {Foj\'on}, \citenamefont {Rivarola},\ and\ \citenamefont
  {Hanssen}}]{Lekadir09}%
  \BibitemOpen
  \bibfield  {author} {\bibinfo {author} {\bibfnamefont {H.}~\bibnamefont
  {Lekadir}}, \bibinfo {author} {\bibfnamefont {I.}~\bibnamefont {Abbas}},
  \bibinfo {author} {\bibfnamefont {C.}~\bibnamefont {Champion}}, \bibinfo
  {author} {\bibfnamefont {O.}~\bibnamefont {Foj\'on}}, \bibinfo {author}
  {\bibfnamefont {R.~D.}\ \bibnamefont {Rivarola}}, \ and\ \bibinfo {author}
  {\bibfnamefont {J.}~\bibnamefont {Hanssen}},\ }\href {\doibase
  10.1103/PhysRevA.79.062710} {\bibfield  {journal} {\bibinfo  {journal} {Phys.
  Rev. A}\ }\textbf {\bibinfo {volume} {79}},\ \bibinfo {pages} {062710}
  (\bibinfo {year} {2009})}\BibitemShut {NoStop}%
\bibitem [{\citenamefont {Galassi}\ \emph {et~al.}(2012)\citenamefont
  {Galassi}, \citenamefont {Champion}, \citenamefont {Weck}, \citenamefont
  {Rivarola}, \citenamefont {Foj\'{o}n},\ and\ \citenamefont
  {Hanssen}}]{Galassi12}%
  \BibitemOpen
  \bibfield  {author} {\bibinfo {author} {\bibfnamefont {M.~E.}\ \bibnamefont
  {Galassi}}, \bibinfo {author} {\bibfnamefont {C.}~\bibnamefont {Champion}},
  \bibinfo {author} {\bibfnamefont {P.~F.}\ \bibnamefont {Weck}}, \bibinfo
  {author} {\bibfnamefont {R.~D.}\ \bibnamefont {Rivarola}}, \bibinfo {author}
  {\bibfnamefont {O.~A.}\ \bibnamefont {Foj\'{o}n}}, \ and\ \bibinfo {author}
  {\bibfnamefont {J.}~\bibnamefont {Hanssen}},\ }\href {\doibase
  10.1088/0031-9155/57/7/2081} {\bibfield  {journal} {\bibinfo  {journal}
  {Phys. Med. Biol.}\ }\textbf {\bibinfo {volume} {57}},\ \bibinfo {pages}
  {2081} (\bibinfo {year} {2012})}\BibitemShut {NoStop}%
\bibitem [{\citenamefont {Tabet}\ \emph {et~al.}(2010)\citenamefont {Tabet},
  \citenamefont {Eden}, \citenamefont {Feil}, \citenamefont {Abdoul-Carime},
  \citenamefont {Farizon}, \citenamefont {Farizon}, \citenamefont {Ouaskit},\
  and\ \citenamefont {M\"ark}}]{Tabet10b}%
  \BibitemOpen
  \bibfield  {author} {\bibinfo {author} {\bibfnamefont {J.}~\bibnamefont
  {Tabet}}, \bibinfo {author} {\bibfnamefont {S.}~\bibnamefont {Eden}},
  \bibinfo {author} {\bibfnamefont {S.}~\bibnamefont {Feil}}, \bibinfo {author}
  {\bibfnamefont {H.}~\bibnamefont {Abdoul-Carime}}, \bibinfo {author}
  {\bibfnamefont {B.}~\bibnamefont {Farizon}}, \bibinfo {author} {\bibfnamefont
  {M.}~\bibnamefont {Farizon}}, \bibinfo {author} {\bibfnamefont
  {S.}~\bibnamefont {Ouaskit}}, \ and\ \bibinfo {author} {\bibfnamefont
  {T.~D.}\ \bibnamefont {M\"ark}},\ }\href {\doibase
  10.1103/PhysRevA.82.022703} {\bibfield  {journal} {\bibinfo  {journal} {Phys.
  Rev. A}\ }\textbf {\bibinfo {volume} {82}},\ \bibinfo {pages} {022703}
  (\bibinfo {year} {2010})}\BibitemShut {NoStop}%
\bibitem [{\citenamefont {Iriki}\ \emph
  {et~al.}(2011{\natexlab{a}})\citenamefont {Iriki}, \citenamefont {Y.},
  \citenamefont {Imai},\ and\ \citenamefont {A.}}]{iriki11a}%
  \BibitemOpen
  \bibfield  {author} {\bibinfo {author} {\bibfnamefont {Y.}~\bibnamefont
  {Iriki}}, \bibinfo {author} {\bibfnamefont {Kikuchi}\ \bibnamefont {Y.}},
  \bibinfo {author} {\bibfnamefont {M.}~\bibnamefont {Imai}}, \ and\ \bibinfo
  {author} {\bibfnamefont {Itoh}\ \bibnamefont {A.}},\ }\href {\doibase
  10.1103/PhysRevA.84.032704} {\bibfield  {journal} {\bibinfo  {journal} {Phys.
  Rev. A}\ }\textbf {\bibinfo {volume} {84}},\ \bibinfo {pages} {032704}
  (\bibinfo {year} {2011}{\natexlab{a}})}\BibitemShut {NoStop}%
\bibitem [{\citenamefont {Iriki}\ \emph
  {et~al.}(2011{\natexlab{b}})\citenamefont {Iriki}, \citenamefont {Y.},
  \citenamefont {Imai},\ and\ \citenamefont {A.}}]{iriki11b}%
  \BibitemOpen
  \bibfield  {author} {\bibinfo {author} {\bibfnamefont {Y.}~\bibnamefont
  {Iriki}}, \bibinfo {author} {\bibfnamefont {Kikuchi}\ \bibnamefont {Y.}},
  \bibinfo {author} {\bibfnamefont {M.}~\bibnamefont {Imai}}, \ and\ \bibinfo
  {author} {\bibfnamefont {Itoh}\ \bibnamefont {A.}},\ }\href {\doibase
  10.1103/PhysRevA.84.052719} {\bibfield  {journal} {\bibinfo  {journal} {Phys.
  Rev. A}\ }\textbf {\bibinfo {volume} {84}},\ \bibinfo {pages} {052719}
  (\bibinfo {year} {2011}{\natexlab{b}})}\BibitemShut {NoStop}%
\bibitem [{\citenamefont {Paredes}\ \emph {et~al.}(2015)\citenamefont
  {Paredes}, \citenamefont {Illescas},\ and\ \citenamefont
  {M\'endez}}]{paredes15}%
  \BibitemOpen
  \bibfield  {author} {\bibinfo {author} {\bibfnamefont {S.}~\bibnamefont
  {Paredes}}, \bibinfo {author} {\bibfnamefont {C.}~\bibnamefont {Illescas}}, \
  and\ \bibinfo {author} {\bibfnamefont {L.}~\bibnamefont {M\'endez}},\ }\href
  {\doibase 10.1140/epjd/e2015-60106-8} {\bibfield  {journal} {\bibinfo
  {journal} {Eur. Phys. J. D}\ }\textbf {\bibinfo {volume} {69}},\ \bibinfo
  {pages} {178} (\bibinfo {year} {2015})}\BibitemShut {NoStop}%
\bibitem [{\citenamefont {Guly\'as}\ \emph {et~al.}(2012)\citenamefont
  {Guly\'as}, \citenamefont {Igarashi},\ and\ \citenamefont
  {Kirchner}}]{gulyas12}%
  \BibitemOpen
  \bibfield  {author} {\bibinfo {author} {\bibfnamefont {L.}~\bibnamefont
  {Guly\'as}}, \bibinfo {author} {\bibfnamefont {A.}~\bibnamefont {Igarashi}},
  \ and\ \bibinfo {author} {\bibfnamefont {T.}~\bibnamefont {Kirchner}},\
  }\href {\doibase 10.1088/0953-4075/45/8/085205} {\bibfield  {journal}
  {\bibinfo  {journal} {J. Phys. B}\ }\textbf {\bibinfo {volume} {45}},\
  \bibinfo {pages} {085205} (\bibinfo {year} {2012})}\BibitemShut {NoStop}%
\end{thebibliography}%

\end{document}